\documentstyle {article}

\begin{document}

{\bf The Little-Parks effect in an inhomogeneous superconducting ring}

\

A.V.Nikulov and I.N.Zhilyaev$^{*}$

Institute of Microelectronics Technology and High Purity
Materials, Russian Academy of Sciences, 142432 Chernogolovka, Moscow
Region, Russia.

\

\begin{abstract}
An inhomogeneous superconducting ring (hollow cylinder) placed in a
magnetic field is considered. It is shown that the superconducting
transition of the section with the lowest critical temperature may be a
first order phase transition if the magnetic flux contained within the ring
is not divisible by the flux quantum. In the vicinity of this transition,
thermal fluctuations can induce the voltage in the ring with rather small
sizes.

PACS number: 74.40. +k \end{abstract}

\vspace{0.3in}

\section{Introduction}

	Superconductivity is a macroscopic quantum phenomenon. One of the
consequences of this fact is the Little-Parks effect \cite{little} which has
been explained by M.Tinkham \cite{tinkham}. The experiment by Little and
Parks is one of the first manifestations of the macroscopic quantum nature
of superconductivity considered in many textbooks (see for example
\cite{tink75}). Little and Parks  discovered that the critical temperature,
$T_{c}$, of a superconducting tube with narrow wall depends in a  periodic
way on a magnetic flux value within the tube. This effect is a consequence
of the dependence of the velocity of the superconducting electrons along
the tube circumference on the magnetic flux value.

	The velocity of the superconducting electrons $v_{s}$ is given by
(See ref.\cite{tink75})

$$v_{s} = \frac{1}{m}(\hbar \frac{d\phi}{dr} - \frac{2e}{c}A) =
\frac{2e}{mc}(\frac{\Phi_{0}}{2\pi }\frac{d\phi}{dr} - A) \eqno{(1)}$$
where $\phi$ is the phase of the wave function $\Psi = |\Psi | \exp(i\phi)$
of the superconducting electrons; $\Phi_{0} = \pi \hbar c/e$ is the flux
quantum; A is the vector potential; m is the electron mass and e is the
electron charge. As a consequence of this relation the velocity along the
tube (or ring) circumference must have fixed values dependent on the
magnetic flux because

$$\int_{l} dl v_{s} = \frac{2e}{mc}(\Phi_{0}n - \Phi) \eqno{(2)}$$
and $n = \int_{l} dl(1/2\pi )d\phi/dr$ must be an integer number since the
wave function must be a simple function. Here  $l=2\pi R$ is the tube (or
ring) circumference; R is the tube (ring) radius; $\Phi = \int_{l}dlA$ is
the magnetic flux contained within the ring. If the tube (ring) wall, w, is
narrow enough (i.e. if $w \ll \lambda$), one has $\Phi \simeq BS$, because
in this case  the magnetic field induced by the superconducting current in
the tube is small.  Here B is the magnetic induction induced by an
external magnet,
$S=\pi R^{2}$ is the area of the tube cross-section and $\lambda$ is the
penetration depth of the magnetic field.

	The energy of the superconductor  increases with the superconducting
electron velocity. Therefore the $|v_{s}|$ tends towards a minimum
possible value. If $\Phi/\Phi_{0}$ is an integer number, this value is equal
to zero. But $v_{s}$ cannot be equal to zero if $\Phi/\Phi_{0}$ is not an
integer number. Consequently, the energy of the superconducting state of
the tube depends in a periodic manner on the magnetic field value. It is the
cause of the Little-Parks effect.

	If the tube (the ring) is homogeneous, then $\int_{l} dl v_{s} = 2\pi
Rv_{s}$. This case was considered in reference \cite{tinkham}. According to
ref.\cite{tinkham},  the critical temperature of the ring is shifted
periodically in the magnetic field:

$$T_{c}(\Phi) = T_{c}[1 - (\xi(0)/R)^{2}(n-\Phi /\Phi_{0})^{2}]
\eqno{(3)} $$ because the $v_{s}^{2}$ value changes periodically with
magnetic field. Here $\xi(0)$ is the coherence length at T = 0. The value
of  ($n-\Phi /\Phi_{0}$)  changes from -0.5 to 0.5. The $T_{c}$ shift is
visible if the tube radius is small enough (if R is little more than
$\xi(0)$.

	In the present work an inhomogeneous ring (tube) with the narrow wall (the
wall thickness $w \ll R, \lambda$) is considered. (In a ring the
height h is smaller than the radius R (i.e. $h < R$) and in the tube $h >
R$). Below the denomination "ring" is used in both cases.
We consider a ring whose  critical temperature varies
along the circumference $l = 2\pi R$, but is constant along the height h.
In such a ring, the magnetic flux shifts the critical temperature of a
section with a lowest $T_{c}$ value only. When the
superconducting state is closed in the ring, the current of the
superconducting electrons must appear as a consequence of the relation (2)
if
$\Phi/\Phi_{0}$ is not an integer number. Therefore the lowest $T_{c}$
value will be shifted periodically in the magnetic field as well as $T_{c}$
of the homogeneous ring. But in addition, the superconducting transition of
the section with the lowest
$T_{c}$ may be a first order phase transition under some conditions. This
principal feature of the inhomogeneous ring is considered in the present
work.

\section{Mean field approximation}

	Let us consider a ring consisting of two sections $l_{a}$ and $l_{b}$
($l_{a}+l_{b} = l =2\pi R$) with different values of the critical
temperature $T_{ca} > T_{cb}$. According to the relation (2) the
superconducting current along the ring circumference, $I_{s}$, must appear
below $T_{cb}$ if $\Phi/\Phi_{0}$ is not an integer number. It is obvious
that the current value must be equal in both sections. Therefore if the
current of the normal electrons is absent, the value of the superconductor
currents $I_{sa}$ and $I_{sb}$ must be equal to

$$I_{s}=I_{sa}=s_{a}j_{sa}=s_{a}2en_{sa}v_{sa}=I_{sb}=s_{b}j_{sb}=
s_{b}2en_{sb}v_{sb} \eqno{(4)} $$
where $n_{sa}$ and $n_{sb}$ are the densities of the superconducting pair
in the sections $l_{a}$ and $l_{b}$;  $v_{sa}$ and $v_{sb}$  are the
velocities of the superconducting pairs in the sections $l_{a}$ and
$l_{b}$ and $s_{a}$ and $s_{b}$ are the areas of wall section of
$l_{a}$ and $l_{b}$. We consider the ring with identical areas $s = s_{a} =
s_{b} = wh$.  $\int_{l} dl v_{s} = v_{sa}l_{a} + v_{sb}l_{b}$. Therefore
according to (2) and (4)

$$v_{sa} = \frac{2e}{mc}\frac{n_{sb}}{(l_{a}n_{sb}+l_{b}n_{sa})}
(\Phi_{o}n-\Phi); \ v_{sb} = \frac{2e}{mc}
\frac{n_{sa}}{(l_{a}n_{sb}+l_{b}n_{sa})} (\Phi_{o}n-\Phi) \eqno{(5)}$$

Consequently the $v_{sb}$ value decreases with the $n_{sb}$ value
increasing. Therefore the dependence of the energy of the superconducting
state on the $n_{sb}$ value can have a maximum in some temperature region
at $T \simeq T_{cb}(\Phi)$. The presence of such a maximum means that the
superconducting transition is a first order phase transition.

	The existence of the maximum and the width of the temperature region
where the maximum exists depends on the $n-\Phi/\Phi_{o}$ value and on the
ring parameters: $l_{a}$, $l_{b}$, w, h and $T_{ca}/T_{cb}$. We
consider this dependence in the present paper. It is shown that these
dependencies can be reduced to two parameters, $B_{f}$ and $L_{I}$, which
are introduced below. It is obvious that the maximum can exist at only
$n-\Phi/\Phi_{0} \neq 0$. Therefore only this case is considered
below.

	The hysteresis of the superconducting transition can be observed if the
maximum is high enough. The maximum height is determined by a parameter F,
which is introduced below. The hysteresis will be observed if the maximum
height is much greater than the energy of the thermal fluctuation, $k_{B}T$.
In the opposite case the thermal fluctuation switches the $l_{b}$
section from the normal state into the superconducting one and backwards
at $T \simeq T_{cb}(\Phi)$. This case is also considered below. It is shown
that the switching between the normal and the superconducting states induces
a voltage on the $l_{b}$ section.

	The Ginsburg-Landau free energy of the ring can be written as

$$F_{GL} = s[l_{a}((\alpha_{a}+\frac{mv_{sa}^{2}}{2})n_{sa}+
\frac{\beta_{a}}{2}n_{sa}^{2})
+ l_{b}((\alpha_{b}+\frac{mv_{sb}^{2}}{2})n_{sb}+ \frac{\beta_{b}}{2}
n_{sb}^{2})] + \frac{LI_{s}^{2}}{2} \eqno{(6)}$$
Here L is the inductance of the ring. $\alpha_{a} =
\alpha_{a0}(T/T_{ca}-1)$,
$\beta_{a}$, $\alpha_{b} = \alpha_{b0}(T/T_{cb}-1)$ and $\beta_{b}$ are
the coefficients of the Ginsburg-Landau theory. We do not consider the
energy connected with the density gradient of the superconducting
pair. It can be shown that this does not influence essentially the results
obtained below.

	The Ginsburg-Landau free energy (6) consists of $F_{GL,la}$ (the energy
of the section $l_{a}$), $F_{GL,lb}$ (the energy of the section $l_{a}$)
and $F_{L}$ (the energy of the magnetic field induced by the
superconducting current):

$$F_{GL} = F_{GL,la} + F_{GL,lb} + F_{L} \eqno{(7)}$$

Substituting the relation (4) for the superconducting current and the
relation (5) for the velocity of the superconducting electrons into the
relation (6), we obtain

$$F_{GL,la} = sl_{a}(\alpha_{a}(\Phi ,n_{sa},n_{sb})n_{sa} +
\frac{\beta_{a}}{2}n_{sa}^{2}) \eqno{(7a)}$$

$$F_{GL,lb} = sl_{b}(\alpha_{b}(\Phi
,n_{sa},n_{sb})n_{sb}+\frac{\beta_{b}}{2} n_{sb}^{2}) \eqno{(7b)}$$

$$F_{L} = \frac{2Ls^{2}e^{2}}{mc} \frac{(\Phi_{0}n-
\Phi)^{2}n_{sa}^{2}n_{sb}^{2}} {(l_{a}n_{sb}+l_{b}n_{sa})^{2}}
\eqno{(7c)}$$
Here $$\alpha_{a}(\Phi ,n_{sa},n_{sb}) = \alpha_{a0} (\frac{T}{T_{ca}}-1 +
(2\pi \xi_{a}(0))^{2} \frac{(n-\Phi /\Phi_{0})^{2}n_{sb}^{2}}
{(l_{a}n_{sb}+l_{b}n_{sa})^{2}})$$
$$\alpha_{b}(\Phi ,n_{sa},n_{sb}) =
\alpha_{b0}(\frac{T} {T_{cb}}-1 +(2\pi \xi_{b}(0))^{2} \frac{(n-\Phi
/\Phi_{0})^{2}n_{sa}^{2}}{(l_{a}n_{sb}+l_{b}n_{sa})^{2}} ) $$ $\xi_{a}(0) =
(\hbar^{2}/2m\alpha_{a0})^{1/2}$; $\xi_{b}(0) =
(\hbar^{2}/2m\alpha_{b0})^{1/2}$ are the coherence lengths at T=0.

	According to the mean field approximation the transition into the
superconducting state of the section $l_{b}$ occurs at $\alpha_{b}(\Phi
,n_{sa},n_{sb}) = 0$. Because $n_{sa} \neq 0$ at $T = T_{cb}$ the
position of the superconducting transition of the $l_{b}$ section depends
on the magnetic flux value:

$$T_{cb}(\Phi) = T_{cb}[1 - (2\pi \xi_{b}(0))^{2} \frac{(n-\Phi
/\Phi_{0})^{2}n_{sa}^{2}} {(l_{a}n_{sb}+l_{b}n_{sa})^{2}}] \eqno{(7d)} $$.

At $l_{a} = 0$ the relation (7d) coincides with the relation (3) for a
homogeneous ring. A similar result ought be expected at $l_{b} \gg
l_{a}$.  But at $l_{b} \ll l_{a}$ the $T_{cb}(\Phi)$ value depends strongly
on the $n_{sb}$ value. At $n_{sb} = 0$ $T_{cb}(\Phi) = T_{cb}[1 - (2\pi
\xi_{b}(0)/l_{b})^{2}(n-\Phi /\Phi_{0})^{2}]$ whereas at $l_{a}n_{sb} \gg
l_{b}n_{sa}$ $T_{cb}(\Phi) = T_{cb}[1 - (2\pi \xi_{b}(0)/l_{a})^{2}(n-\Phi
/\Phi_{0})^{2}]$. Consequently a hysteresis of the superconducting
transition ought be expected in a ring for $l_{b} \ll l_{a}$.

	To estimate the dependence of the hysteresis value on the ring
parameters, we transform the relation (7) using the relations for the
thermodynamic critical field $H_{c}=\Phi_{0}/2^{3/2}\pi \lambda_{L}\xi$;
$\alpha^{2}/2\beta = H_{c}^{2}/8\pi $ and for the London penetration depth
$\lambda_{L}=(cm/4e^{2}n_{s})^{1/2}$. We consider a ring with
$l_{a} \gg \xi_{a}(T) = \xi_{a}(0)(1-T/T_{ca})^{0.5}$. $n_{sa} \simeq
-\alpha_{a}/\beta_{a}$ in this case. Then

$$F_{GL}= F_{GLa} + Fn'_{sb}(\tau + \frac{1}{(n'_{sb}+1)^{2}} +
n'_{sb}(B+\frac{1}{(n'_{sb}+1)^ {2}} (2 +L_{I}))) \eqno{(8)} $$
Here $n'_{sb} = l_{a}n_{sb}/l_{b}n_{sa}$; $$F_{GLa} =
-sl_{a}\frac{H_{ca}^{2}}{8\pi }(1+\frac{(2\pi \xi_{a}(T))^{4}}{l_{a}^{4}}
(\frac{(n-\Phi /\Phi_{0})n'_{sb}}{(n'_{sb}+1)})^{4})$$ Because
$l_{a} \gg 2\pi \xi_{a}$, $F_{GLa} \simeq -sl_{a}H_{ca}^{2}/8\pi $.
$$F = \frac{s\xi_{a}(T)H_{ca}^{2}}{2} \frac{2\pi \xi_{a}(T)}{l_{a}}
(n-\Phi/\Phi_{0})^{2}$$ $$\tau = (\frac{T}{T_{cb}}-1)(n-\Phi
/\Phi_{0})^{-2} \frac{l_{b}^{2}}{(2\pi \xi_{b}(0))^{2}}$$ $$B_{f} =
0.5\frac{\beta_{b}}{\beta_{a}} \frac{l_{b}}{l_{a}} \frac{l_{b}^{2}}{(2\pi
\xi_{b}(0))^{2}} (n-\Phi /\Phi_{0})^{-2}$$ $$L_{I} = 4\pi
\frac{s}{\lambda_{La}^{2}} \frac{L}{l_{a}}$$ For $h > R$, $L= k4\pi
R^{2}/h$ where k = 1 at $h \gg R$. Consequently, $L_{I} = 4\pi
(l/l_{a})(lw/\lambda_{La}^{2}(T))$ in this case. At $h, w \ll R$, $L
\simeq 4l\ln(2R/w)$, therefore $L_{I} = 16\pi
(l/l_{a})(s/\lambda_{La}^{2}(T))\ln(2R/w)$ in this case.

	The numerical calculations show that the $F_{GL}(n'_{sb})$ dependence
(8) has a maximum at small enough values of $B_{f}$ and $\L_{I}$ in some
region of the $\tau$ values. The width of the $\tau$ region with the
$F_{GL}(n'_{sb})$ maximum depends on the $B_{f}$ value first of all.
At $\L_{I} \ll 2$ the maximum exists at $B_{f} < 0.4$. For example at
$B_{f} = 0.2$ and $L_{I} \ll 2$ the maximum takes place at $-1.02 < \tau <
-0.89$.  This means that the transition into the superconducting state of
the section $l_{b}$ occurs at $\tau \simeq -1.02$, (that is at
$T_{cs} = T_{cb}(1 - 1.02(n-\Phi /\Phi_{0})^{2} (2\pi
\xi_{b}(0)/l_{b})^{2}$) and the transition in the normal state occurs at
$\tau \simeq -0.89$, (that is at $T_{cn} = T_{cb}(1 - 0.89(n-\Phi
/\Phi_{0})^{2} (2\pi \xi_{b}(0)/l_{b})^{2}$) if  thermal fluctuations are
not taken into account.

	The inequality $\L_{I} \ll 2$ is valid for a tube (when $h > R$) with
$2\pi lw \ll \lambda_{La}^{2}(T)$ and for a ring (when $h < R$) with $8\pi
hw \ll \lambda_{La}^{2}(T)$. The hysteresis value increases with decreasing
$B_{f}$ value and decreases with increasing $B_{f}$ value.
The $B_{f}$ value is proportional to $(n-\Phi /\Phi_{0})^{-2}$.
Consequently, the hysteresis value depends on the magnetic field value.
Because the hysteresis is absent at $B_{f} > 0.4$, it can be observed in
the regions of the magnetic field values, where $\Phi/\Phi_{0}$ differs
essentially from an integer number. The width of these regions depends on
the $0.5(\beta_{b}/\beta_{a}) (l_{b}^{3}/(2\pi \xi_{b}(0))^{2}l_{a})$ value
(see above the relation for $B_{f}$). Since $(n-\Phi /\Phi_{0})^{2} <
0.25$ and $\beta_{b} \simeq \beta_{a}$ in the real case, the hysteresis
can be observed in the ring with $l_{b}^{3} < 0.2(2\pi
\xi_{b}(0))^{2}l_{a})$. For example in the ring with $l_{b} = 2\pi
\xi_{b}(0)$ and $l_{a} = 10l_{b}$, the hysteresis can be observed at
$|n-\Phi /\Phi_{0}| > 0.35$ (if $\beta_{a} = \beta_{b}$). At $|n-\Phi
/\Phi_{0}| = 0.5$  $B_{f} = 0.2$ and the hysteresis is equal to $T_{cn} -
T_{cs} \simeq 0.03T_{cb}$ in this ring.

\section{Thermal fluctuation theory}

	Above we have used the mean field approximation which is valid when the
thermal fluctuation is small. In our case the mean field approximation is
valid if the height of the $F_{GL}-F_{GLa}$ maximum, $F_{GL,max}$, is much
greater than $k_{B}T$. This height depends on the F, $\tau$, $B_{f}$ and
$L_{I}$ values: $F_{GL,max} = FH(\tau ,B_{f},L_{I})$. The F parameter is
determined above. The $H(\tau ,B_{f},L_{I})$ dependence can be
calculated numerically from the relation (8). To estimate the validity of
the mean field approximation we ought to know the maximum value of the
$H(\tau )$ dependence: $H_{max}(B_{f},L_{I})$. We can use
the mean field approximation if $FH_{max}(B_{f},L_{I}) \gg k_{B}T$. This is
possible if the height of the ring is large enough, namely

$$h \gg \xi_{a}(0)\frac{1}{\pi H_{max}(B_{f},L_{I})}\frac{l_{a}}{w}
\frac{Gi^{1/2}} {T_{ca}/T_{cb} -1}(n- \frac{\Phi}{\Phi_{0}})^{-2}$$
Here $Gi = (k_{B}T_{ca}/\xi_{a}(0)^{3}H_{ca}^{2})^{2}$ is the Ginsburg
number of a three-dimensional superconductor. We have used the relation
for the F parameter (see above). For conventional superconductors $Gi =
10^{-11} - 10^{-5}$. $H_{max} \simeq 10^{-2}$ for typical $B_{f}$
and $L_{I}$ values. For example in the ring with $B_{f} = 0.2$ and $L_{I}
\ll 2$ the $H(\tau )$ dependence has a maximum $H_{max}(B_{f} = 0.2,L_{I}
\ll 2) = 0.024$ at $\tau \simeq -0.94$. Consequently, the  value of h
cannot be very large. As an example for a ring with
parameter value
$B_{f} = 0.2$,
$L_{I} \ll 2$, $l_{a}/w = 20$, $T_{ca}/T_{cb} - 1 = 0.2$, and fabricated
from an extremely dirty superconductor with $Gi = 10^{-5}$, the mean field
approximation is valid at $h \gg 20 \xi_{a}(0)$ if $|n-\Phi /\Phi_{0}|
\simeq 0.5$.

	If the mean field approximation is not valid, we must take into account
the thermal fluctuations which decrease the value of the
hysteresis. The probability of the transition from normal into
superconducting state and that of the transition from superconducting
into normal state are large when the maximum value of
$F_{GL}-F_{GLa}$ is no much more than $k_{B}T$.  Therefore the hysteresis
can not be observed at $FH_{max}(B_{f},L_{I}) < k_{B}T$. This inequality
can be valid for a ring made by lithography and etching methods from a thin
superconducting film, where h is the film thickness in such a ring.

	Let the $l_{b}$ section of this ring have the lowest critical
temperature $T_{cb}$. As a consequence of the thermal fluctuations, the
density $n_{sb}(r,t)$ changes with time. We can consider $n_{sb}(r,t)$ as a
function of the time only if $h,w,l_{b} \simeq \ or \ < \xi_{b}(T)$. At $T
\simeq T_{cb}(\Phi)$ (at the resistive transition) $l_{b}$ is switched
by the fluctuations from the normal state in the superconducting state and
backwards i.e. some times ($\simeq t_{n}$) $n_{sb} =0$ and some times
($\simeq t_{s}$) $n_{sb} \neq 0$. According to (5) $v_{sb}$
can not be equal to zero if $n_{sa} \neq 0$ and $\Phi_{o}n-\Phi \neq 0$.
Therefore the superconducting current value $I_{s}$ (see (4)) changes also
with time.

	The change of the superconducting current induces the change of the
magnetic flux $\Phi = H\pi R^{2} + L(I_{s}+I_{n})$ and, as a consequence,
induces the voltage and the current of the normal electrons (the normal
current, $I_{n}$). The total current $I=I_{s}+I_{n}$ must be equal in
both sections, because the capacitance is small. But $I_{sa}$ cannot be
equal to $I_{sb}$. Then $I_{na} \neq I_{nb}$ in this case and consequently,
the potential difference $dU/dl$ exists along the ring circumference. Then
the electric field along the ring circumference $E(r)$ is equal to

$$E(r) = -\frac{dU}{dl} -\frac{1}{l}\frac{d\Phi}{dt} =-\frac{dU}{dl} -
\frac{L}{l}\frac{dI}{dt}= \frac{\rho_{n}}{s} I_{n} \eqno{(9)}$$
where $\rho_{n}$ is the normal resistivity. If the time of the normal state
is much smaller than the decay time of the normal current, $t_{n} \ll
L/R_{nb}$, the total current I is approximately constant in time. $R_{bn} =
\rho_{bn}l_{b}/s$ is the resistance of the section $l_{b}$ in the normal
state. If also $t_{s} \gg L/R_{bn}$, then

$$I \simeq s2en_{sa}<v_{sa}> \simeq s\frac{4e^{2}}{mc}\frac{n_{sa}<n_{sb}>}
{l_{b}n_{sa}+l_{a}<n_{sb}>}(\Phi_{0}n-\Phi) \eqno{(10)}$$
Here $<n_{sb}>$ is the thermodynamic average of $n_{sb}$. If $t_{n} \neq 0$,
the resistivity of the $l_{b}$ section, $\rho_{b} \simeq
\rho_{bn}t_{n}/(t_{s}+t_{n})$), is not equal to zero. Consequently the
direct potential difference $U_{b}$ can be observed on the section $l_{b}$
at $T \simeq T_{cb}(\Phi)$

$$U_{b} = R_{b}I \simeq \frac{l_{b}<n_{sb}>} {l_{b}n_{sa}+l_{a}<n_{sb}>}
\frac{(\Phi_{0}n-\Phi)}{\lambda_{La}^{2}}\rho_{b} \eqno{(11)}$$

According to (11) $U_{b} \neq 0$ at $<n_{sb}> \neq 0$ and $\rho_{b} \neq
0$. Consequently, the direct potential difference can be observed in the
region of the resistive transition  of the section $l_{b}$ (where $0 <
\rho_{b} < \rho_{bn}$).

	The relation (11) is valid at $t_{n} \ll L/R_{bn}$. At $t_{n} \gg
L/R_{bn}$ the direct part of the potential difference is equal to

$$U_{b} \simeq \frac{s<n_{sb}>} {l_{b}n_{sa}+l_{a}<n_{sb}>}
\frac{(\Phi_{0}n-\Phi)}{\lambda_{La}^{2}}Lf \eqno{(12)}$$
where f is the frequency of the switching from the normal into the
superconducting state.

	The Ginsburg-Landau free energy $F_{GL}$ changes in time with amplitude
$k_{B}T$ as a consequence of the thermal fluctuation. According to the
relations (10),(12) and (7c) $U_{b}I/f \simeq F_{L}$. Because $F_{L}$ is a
part of $F_{GL}$ (see (7)) the $U_{b}$ value can not exceed
$(k_{B}TR_{bn}f)^{1/2}$. The maximum value of the switching frequency f is
determined by the characteristic relaxation time of the superconducting
fluctuation $\tau_{GL}$:  $f_{max} \simeq 1/\tau_{GL}$. In the linear
approximation region \cite{skocpol}

$$\tau_{GL} = \frac{\hbar}{8k_{B}(T-T_{c})} \eqno{(13)}$$

The width of the resistive transition of the section $l_{b}$ can be
estimated by the value $T_{cb}Gi_{b}$. $Gi_{b} = (k_{B}T/H_{c}^{2}(0)l_{b}
s)^{1/2}$ is the Ginsburg number of the section $l_{b}$. Consequently the
$U_{b}$ value can not be larger than

$$U_{b,max} = (\frac{8R_{b}Gi_{b}}{\hbar})^{1/2}k_{B}T_{cb} \eqno{(14)}$$

The $U_{b,max}$ value is large enough to be measured
experimentally. Even at $T_{c} = 1 \ K$ and for real values $R_{b} = 10
\Omega$ and $Gi_{b} = 0.05$, the maximum voltage is equal to $U_{b,max}
\simeq 3 \mu V$. In a ring made of a high-Tc superconductor, $U_{b,max}$
can exceed $100 \mu V$. One ought to expect that the real $U_{b}$ value
will be appreciably smaller than $U_{b,max}$. This voltage can be determined
 by the periodical dependence on the magnetic field value (see the
relations (11) and (12)).

\section*{ACKNOWLEDGMENTS} A.V.N. thanks the National Scientific Council on
"Superconductivity" of SSTP "ADPCM" (Project 95040) and INTAS (Project
96-0452) for financial support. I.N.Zh. thanks the Norway Research Council,
Russian Program Solid State Nanostructures (Project 97-1060) and RFBR.

\

$^{*}$ E-mail: zhilyaev@ipmt-hpm.ac.ru

\

\end{document}